# Engineering disorder in three-dimensional photonic crystals

Rajesh V. Nair* and B. N. Jagatap

Atomic and Molecular Physics Division, Bhabha Atomic Research Centre, Mumbai 400 085 India

*Email: rvnair@barc.gov.in

## Abstract

We demonstrate the effect of introducing controlled disorder in self-assembled three-dimensional photonic crystals. Disorders are induced through controlling the self-assembling process using an electrolyte of specific concentrations. Structural characterization reveals increase in disorder with increase in concentrations of the electrolyte. Reflectivity and transmittance spectra are measured to probe the photonic stop gap at different levels of disorder. With increase in disorder the stop gap is vanished and that results in a fully random photonic nanostructure where the diffuse scattered intensity reaches up to 100%. Our random photonic nanostructure is unique in which all scatters have the same size and shape. We also observe the resonant characteristics in the multiple scattering of light.

## I. INTRODUCTION

Light-matter interactions in photonic nanostructures have paramount importance in basic science and technological developments [1]. In this context, light propagation in both ordered and disordered photonic nanostructures has received considerable attention in recent years [2, 3]. In ordered photonic nanostructures, also called photonic crystals, light propagation is forbidden for certain frequency range called the photonic band gap [4]. The basis of photonic band gap is the Bragg diffraction along different crystallographic directions and polarizations [5]. In the frequency range of photonic band gap, photon density of states goes zero and that results into an ultimate control on the light emission and propagation [1, 4]. On the other extreme, in disordered photonic nanostructures, light undergoes multiple scattering and the light propagation can be approximated as a diffusion process. Further these structures can exhibit dramatic optical effects such as the Anderson localization of light [6], coherent backscattering of light [7], random lasing [8] and focusing of light [9]. The conventional disordered photonic nanostructures documented in literature [6, 8-10] consists of polydisperse scatters, which result in the inhibition of any sort of resonant effects associated with multiple scattering of light [11]. All photonic crystals that are synthesized to date have intrinsic disorders due to size dispersions of the building blocks, lattice displacement and other imperfections that occur during the growth process. While the intrinsic disorder is inevitable in photonic crystals, the goal is to reduce the disorder to minimum. It is, in this context, very important to investigate the dependence of controlled disorder on the optics of photonic crystals. Another interest stems from the fact that photonic crystals with controlled disorder are ideal candidates to realize the Anderson localization of light [12]. It may be stressed here that while scattering of light by intrinsic disorder, which is inevitable in photonic crystals, is a contemporary topic of interest and well documented in literature [13-18], the effect of controlled (extrinsic) disorder induced characteristics of the photonic stop gap is seldom investigated.

In this communication, we report experimental results of controlled disorder induced changes in the photonic stop gap and the scattering strength of photonic nanostructures. We demonstrate here how the structural and optical properties of a self-assembled three-dimensionally (3D) ordered photonic crystals change from the near-perfect ordering to a completely disordered photonic nanostructure. Hence photonic nanostructures between near-perfect and completely random ordering can be achieved. Unlike the conventional disordered photonic nanostructure [6, 8-10], our photonic nanostructure is unique in the sense that each scatters has the same size and shape, and therefore one expects resonant scattering [11]. In addition, each scatters are internally doped with Rhodamine B dye, which opens a platform to probe the modification of emission characteristics from random photonic nanostructures. This paper is organized as follows: The experimental details and methods used for the characterization of photonic nanostructures are given in Sec. II. A detailed discussion of the structural and optical results at different level of disorder in photonic crystals is presented in Sec.

III. The future prospects of our work are briefly outlined. Important conclusions arising from this work are given in Sec. IV.

## II. EXPERIMENTAL DETAILS

We use colloidal suspensions of Rhodamine-B dyed polystyrene (PS-RhB) spheres of diameter 295 nm with a polydispersity index of 0.1. Controlled disorder is introduced by breaking the stability of colloidal suspensions with addition of electrolytes, e.g., $CaCl_2$, prior to growth as proposed in Ref. 19. Colloidal suspensions are generally charge stabilized so that they never cluster in the suspensions, and that prevents the ordering during self-assembly. We reduce the electrostatic repulsive potential between the spheres by adding the electrolyte of appropriate concentrations. In the present case, the surface charge of PS-RhB spheres is negative and therefore positive ions are necessary to screen the potential. We use 200 μL of 2.5 wt% PS-RhB colloidal suspensions together with $CaCl_2$ electrolyte at different concentrations so as to generate controlled amount of disorder in photonic crystals. The colloidal suspension is spread on a clean glass substrate (2.5 cm × 2.5 cm) using a laboratory syringe and allowed to evaporate which results in samples in a time span of three hours. The experimental details may be found elsewhere [20].

Field emission scanning electron microscope (FE-SEM) images are taken for photonic crystals grown with different concentrations of $CaCl_2$ to corroborate structural evolution of the disorder. All the samples are sputtered with silver to avoid the charging during the electron beam bombardment with the samples. Optical characteristics of the photonic nanostructures are probed through the static measurements of reflectivity and transmittance spectra using a Perkin-Elmer spectrophotometer with Halogen lamp as the light source. The light beam is unpoalrized and having a spot size of 5 mm × 5 mm. A silicon detector is used for collecting the reflected or transmitted light. Total transmission spectra are measured using an integrating sphere attached to the spectrophotometer to probe the resonant characteristics associated with multiple scattering of photons.

## III.  RESULTS AND DISCUSSIONS

FE-SEM image of the photonic crystals synthesized using 0M, and $10^{-2}$M $CaCl_2$ electrolyte is shown in Fig. 1(a) and 1(b), respectively. In Fig. 1(a) very good ordering of spheres is evident. Here each sphere is surrounded by six other spheres and represents the (111) plane of the face centered cubic (*fcc*) lattice [20]. In Fig. 1(b), FE-SEM image of the photonic crystal shows the displacement of spheres in addition to the lattice disorder. Short range ordering is still present but the fine structural ordering vanished as compared to Fig. 1(a). Fig. 1(c) and 1(d) show the image of the photonic crystal synthesized using $10^{-1}$M $CaCl_2$ electrolyte. In Fig. 1(c), the complete removal of periodic nature is evident and photonic crystal becomes a random photonic nanostructure. Fig. 1(d) shows the same photonic crystal image (*cf*.Fig.1(c)) in higher magnification. The introduction of disorder is clearly visible and each sphere is connected as no sphere can stand in the air. It is interesting to see the clustering of three spheres due to the addition of higher concentration of electrolyte. The effect of disorder may be more appreciated on the optical properties, which are discussed and analyzed in what follows.

The signature of photonic stop gap is probed through the measurements of reflectivity (R) and transmittance (T) spectra. These for the photonic crystals synthesized using 0M $CaCl_2$ are shown using a black line in Fig. 2 (a) and 2(b) respectively. The peak in R spectra coincides with the trough in T spectra, which is the signature of photonic stop gap [20]. The photonic stop gap is centered at 603 nm with a peak reflectance of 56%. The photonic strength (S) or the full width at half maximum of the reflectivity peak is estimated to be 6.63%. The oscillations in the long wavelength side of the stop gap are the Fabry-Perot (F-P) oscillations and these are due to the uniform thickness of the sample. The thickness (t) of the photonic crystal film is estimated from F-P fringes using the relation [21], t = $1239/2n_{eff}\Delta E$, where $n_{eff}$ is the effective refractive index of the photonic crystal and $\Delta E$ is the difference in energy positions in eV of two consecutive troughs in the F-P oscillations. The thickness is estimated to be 6.2 μm (number of layers ~ 26). Since all the photonic crystals are grown in identical conditions, we expect their thickness and dimensions to be same. The Bragg length ($L_B$), the distance through which light propagate at the stop gap wavelength, is calculated using the formulae [22], $L_B = 2d_{111}/\pi S$, where $d_{111}$ is the distance between the lattice planes along the [111] direction. The estimated value of $L_B$ is 2.3 µm and $t \sim 2.7 L_B$. This indicates that finite size effects are eliminated in photonic crystals and they are strongly interacting with light.

The red line in fig. 2(a) and 2(b) represents the photonic stop gap of the photonic crystal synthesized using $10^{-4}$M $CaCl_2$. The photonic stop gap is very well-resolved, which indicates that the very good ordering is preserved in the depth of the sample. The very low concentration of the electrolyte is not sufficient enough to introduce the disorder. Photonic stop gap of the photonic crystal at an electrolyte concentration of $10^{-3}$M is shown using a blue line in Fig. 2(a) and 2(b). It is evident that this concentration of the electrolyte also does not induce any appreciable disorder in the photonic crystals. On going from 0M to $10^{-3}$M concentration of the electrolyte, the widths of the R and T spectra remain the same. The transmittance value at the stop gap wavelength, however, increases which indicates that the disorders start building up in the structure, since the transmitted beam probes the total thickness of the sample. The higher-order stop gaps [23], which are extremely sensitive to disorder, are present for these photonic crystals (shown using a downward arrow in Fig. 2(a)) and they confirm the minimal amount of disorders. For electrolyte concentrations of $3 \times 10^{-3}$ M and $1 \times 10^{-2}$ M, photonic stop gap is shown using orange and blue lines in Fig. 2(a) and 2(b) respectively. The photonic stop gaps are starting to disappear and the higher-order reflectivity peak also vanished as seen in Fig. 2(a). The peak reflectance at the stop gap wavelength is drastically decreased and the T spectra exhibit a shallow decrease in intensity towards the shorter wavelength region. This shows that an appreciable amount of disorder is induced in the photonic crystals. This is also seen in the FE-SEM image of the photonic crystals (see Fig. 1(b)). The zoomed long-wavelength F-P fringes are shown as an inset in Fig.1 (a). It is seen that the F-P fringes are well-resolved for photonic crystals grown with 0M $CaCl_2$ (black line). But for photonic crystals grown with $3 \times 10^{-3}$ M (orange line) and $10^{-2}$ M (green line) $CaCl_2$ F-P fringes are blue shifted and less pronounced due to the reduction in $n_{eff}$ as a results of controlled vacancies generated in photonic crystals. The F-P fringes for photonic crystals grown with $10^{-1}$ M (pink line) $CaCl_2$ are completely disappeared owing to the complete disordered nature of the photonic crystal. This serves as another confirmation of the growth of controlled disorder in photonic crystals as the F-P fringes are reduced in a systematic way. The trough seen at 555 nm in the T spectra (indicated with a vertical dashed line) is due to the absorption of Rhodamine B dye. R and T spectra for photonic crystal synthesized using $10^{-1}$ M electrolyte is shown using pink lines in Fig. 2(a) and 2(b). We observe here that R and T are completely zero and the stop gap has disappeared. This indicates that the sample becomes completely disordered and there are no lattice planes to build up the Bragg diffraction. The FE-SEM images of the photonic crystals also confirm the high degree of disorder induced in the photonic crystals (see Fig. 1(c) and 1(d)) as compared to the near perfect ordering seen in Fig. 1(a). In Fig. 2(a) and 2(b), for $10^{-1}$ M electrolyte concentration, no light is detected either in the R or in the T spectra. Therefore it is very important to investigate where the light has disappeared.

We thus see that with increase in electrolyte concentration, disorder starts to increase in the structure. Consequently the diffuse multiple scattering (D) of light must also increase inside the photonic crystals. From the measurement of R and T, we estimate D = 1- (R+T). This is valid for wavelength region of no absorption and in the low energy range (a/λ < 1, where "a" is the *fcc* lattice constant). A perfect photonic crystal has R = 100%, T = 0% and D = 0%. Therefore for a completely disordered photonic crystal, we expect D = 100%. Estimated values of D for photonic crystals synthesized using $CaCl_2$ concentrations of 0 M (black line), $10^{-4}$ M (red line), $10^{-3}$ M (blue line), $3 \times 10^{-3}$ M (orange line), $10^{-2}$ M (green line), and $10^{-1}$ M (pink line) are given in Fig. 2(c). For an electrolyte concentration up to $10^{-3}$ M, diffuse scattering of light is increased with decrease in wavelength. At the stop gap wavelength diffuse scattered light drops to very low value but it is not zero. This is due to the residual (intrinsic) disorder present in photonic crystals. The residual disorder is estimated to be ~ 40% for these electrolyte concentrations. For an electrolyte concentration of $3\times10^{-3}$M and $1\times10^{-2}$M, the residual disorder is ~ 84% and diffuse scattered intensity increases towards the shorter wavelength region in addition to the disappearance of stop gap in the spectra as there exists no lattice planes to build the Bragg diffraction. For $10^{-1}$ M electrolyte concentration, the residual disorder is 99% and the photonic nanostructure become completely random with nearly 100% diffuse scattering of light. The results of total transmission measurements done using an integrating sphere on a near-perfect (0% controlled disorder) and completely disordered photonic crystal (100% controlled disorder) is given in Fig. 3(a) and 3(b), respectively. In fig. 3(a) no resonances are seen in the wavelength ranges comparable to the optical diameter of the constituting spheres. This is due to the transformation of each Mie scattering to Bragg scattering because of the periodic nature of the structure. But in fig. 3(b), for wavelengths comparable to the optical diameter of the spheres, resonances are very evident which are due to the excitation of Mie modes in the spheres (vertical dashed lines). This characteristic gives the experimental evidence for the resonant nature in multiple scattering of light. These kinds of resonances are smeared out in conventional random photonic structures due to the polydispersity of the scatters. It is an interesting topic to further understand the resonant light scattering in photonic crystals with moderate disorder and the interaction between the Bragg and Mie modes [25]. Our results open avenues for studying these kinds of resonances and their interaction.

Our disordered photonic nanostructure consists of scatters doped with Rhodamine B dye and hence there exists a possibility of observing the modification of emission characteristics leading to random lasing [10, 26]. Since the scatters in our random photonic structures have the same size and shape and therefore a resonant behavior in random lasing can be expected [24]. Reduction of diffusion constant and back scattering experiments are reported for colloidal photonic crystals with intrinsic disorder [15, 16]. In our photonic crystals, we can control the disorder precisely, and hence diffusion constant and back scattering experiments on our samples provide more information into light scattering in photonic crystals and experiments are being initiated in this direction. Indeed, a recent work related to the effect of controlled disorder in an alloy-type photonic crystal reveals the growing interest in the study of controlled disorder [27]. Therefore our result opens new avenues to study the physics and applications of the disorder induced effects in photonic crystals in addition to the fascinating physical aspects of mono-disperse random photonic nanostructures.

## IV.   CONCLUSIONS

In conclusion, we have shown the effect of introducing controlled disorder in self-assembled photonic crystals by adding electrolyte at specific concentrations. Structural characterization shows that disorders are generated with increase in electrolyte concentration. Electrolyte concentrations of $10^{-1}$M are sufficient to generate a completely random photonic nanostructure. Reflectivity and transmittance measurements indicate the systematic growth of controlled disorder in photonic crystals. Diffuse scattered light estimated from the reflectivity and transmittance measurements is another confirmation of controlled disorder in photonic crystals. Total transmission measurements show the resonant characteristics in the multiple scattering of photons due to the monodispersity of the scatters in our photonic nanostructures.

**ACKNOWLEDGEMENTS**

The authors thank Prof. R. Vijaya and Prof. S. S. Major, IIT Bombay, India for providing us the colloidal suspensions and spectrophotometer used in the present work, respectively. RVN thank BRNS, Govt. of India for Dr. K S Krishnan Research Fellowship.

# Figures

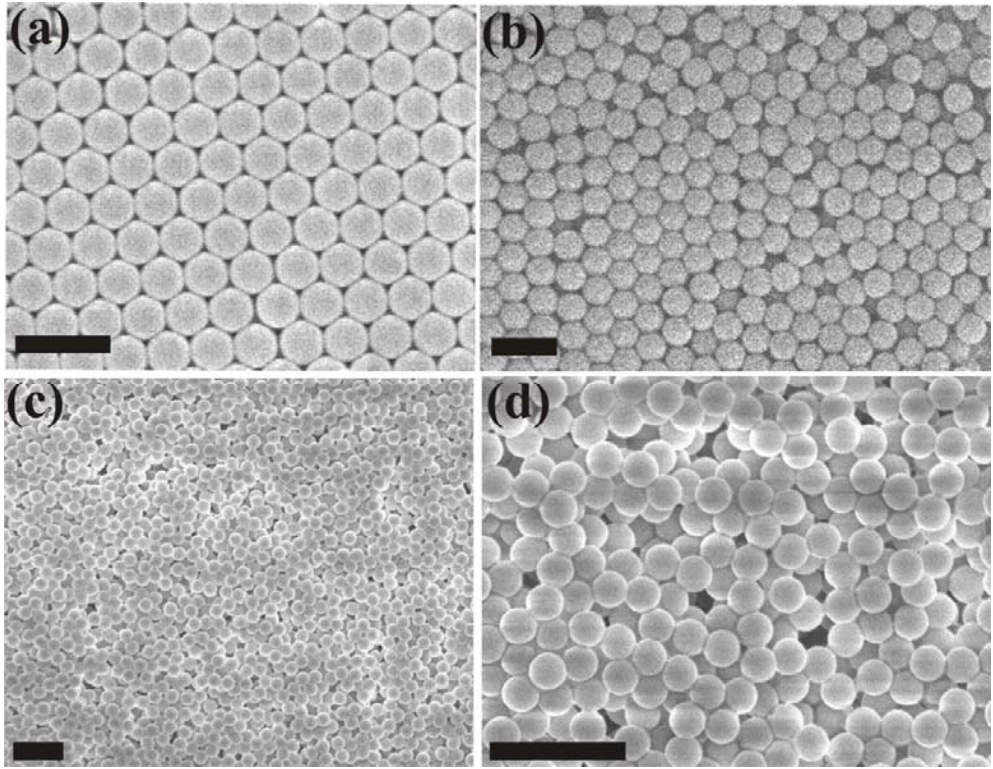

**FIG 1.** FE-SEM images of the photonic crystal structure for an electrolyte concentration of 0M (a), $10^{-2}$ M (b), and $10^{-1}$ M (c and d). The near-perfect ordering of spheres is seen in (a) whereas spheres starts repelling each other results in imperfections in the samples as seen in (b). The complete removal of periodic nature is achieved using $10^{-1}$ M $CaCl_2$ results in a highly disordered photonic nanostructure as seen in (c) and (d). Scale bar is 1 μm.

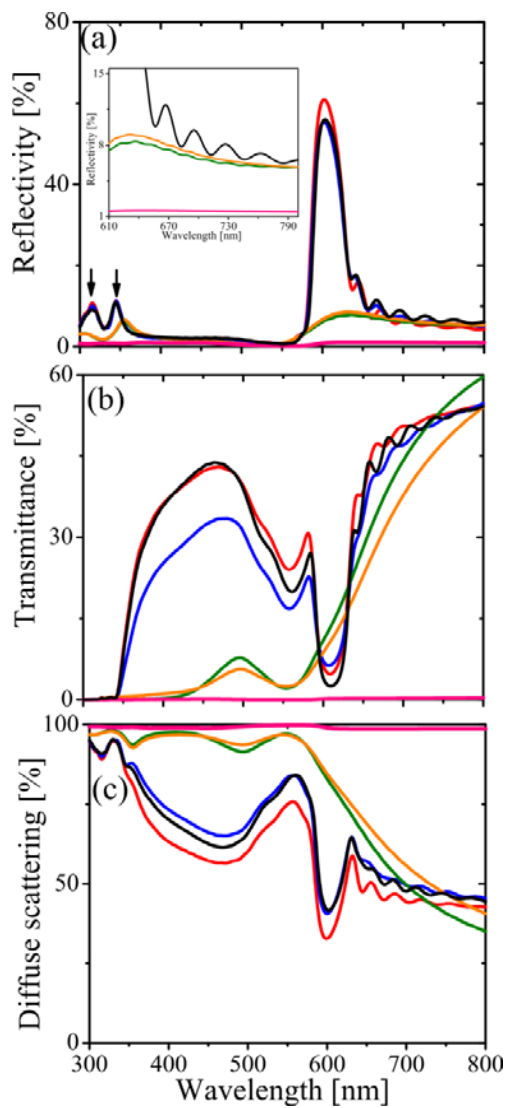

**Fig.2.** [color online] (a) Reflectivity, (b) transmittance and (c) diffuse scattered spectra for photonic crystals synthesized using 0 M $CaCl_2$ (black line), $10^{-4}$ M (red line), $10^{-3}$ M (blue line), $3 \times 10^{-3}$ M (orange line), $10^{-2}$ M (green line), and $10^{-1}$ M (pink line). In (a), with increase in $CaCl_2$ concentrations, peak reflectance reduces and at high concentrations reflectivity goes to zero. Inset gives the long-wavelength Fabry-Perot fringes for photonic crystals grown with different concentrations of electrolyte. In (b), transmittance increases at the stop gap and for higher concentration of the electrolyte, no light is transmitted as seen in (b). The reflectivity peaks seen between 300 to 400 nm in (a) is the higher-order stop gaps (indicated with a downward arrow). The trough seen at 555 nm shown using a vertical red dashed line is due to the absorption of Rhodamine B dye. In (c), the trough seen around 600 nm is due to the stop gap signifies the residual disorder. With increase in $CaCl_2$ concentration, the diffuse scattering of light increases and for $10^{-1}$M concentration the diffuse scattered light is 100%.

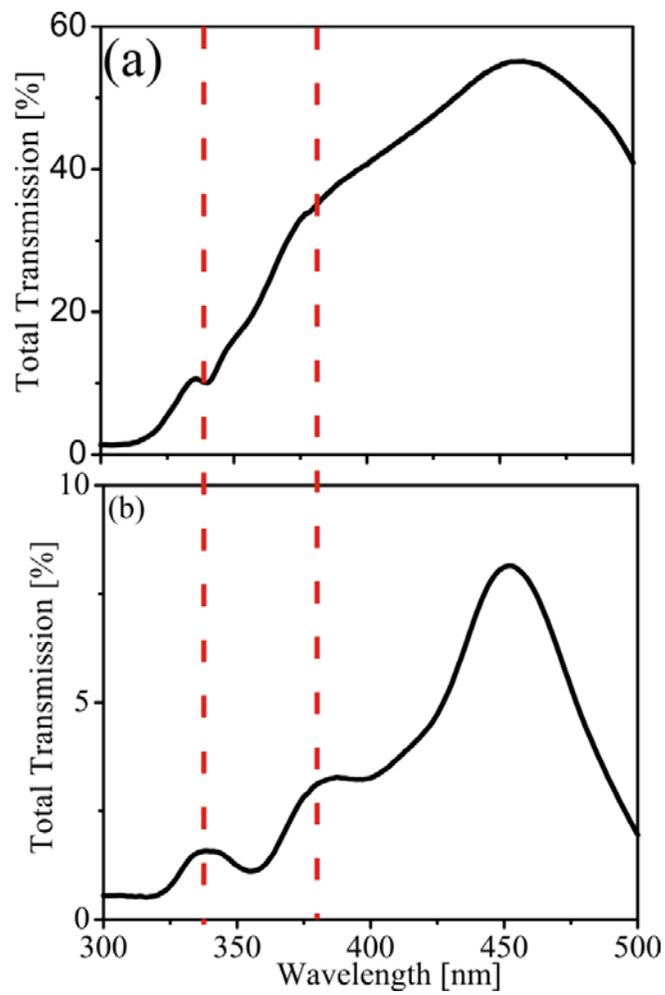

**Fig. 3.** Total transmission spectra for photonic crystal synthesized using (a) 0M $CaCl_2$ and (b) $10^{-1}$M $CaCl_2$. Resonances are evident at the wavelength ranges comparable to the optical diameter of the spheres due to the excitation of Mie modes in (b) whereas those resonances are absent in (a). These kinds of resonances are signature of resonant nature in the multiple scattering of light.